\documentstyle[11pt,newpasp,twoside,epsfig]{article}
\reversemarginpar

\begin{document}
\title{Evolution from OB Associations and Moving Groups to the Field
Population}
\author{Anthony G.A.\ Brown}
\affil{European Southern Observatory, Karl-Schwarzschildstra{\ss}e 2, D-85748
Garching bei M\"unchen, Germany}

\begin{abstract}
If we want to understand the evolution from star clusters to the Galactic
field star population, the Solar neighbourhood is an ideal place to start. It
contains objects from dense, very young clusters to old and almost dissolved
moving groups. I describe the observational evidence for the ongoing
dissolution of OB associations and open clusters. Subsequently, the
interpretation of the local phase space distribution of stars in terms of old
moving groups is discussed, emphasising the present limitations. Finally, I
discuss future observational requirements and the possibilities offered by
upcoming astrometric space missions.
\end{abstract}

\section{Introduction}
\label{sec:intro}

The basic motivation for studying the origin of field stars is the desire to
understand the formation history of the entire Galaxy. In the case of the
Galactic disk we wish to understand its abundance and age patterns by studying
where the field stars were born and how they migrate through the disk. This
will eventually lead to constraints on, e.g., the homogeneity of the Galactic
interstellar medium. In the case of the Halo we want to answer questions about
the degree to which it has been built up by Dwarf galaxies merging with our
own galaxy and how this affects the Halo's phase space structure.

On smaller scales we wish to understand what fraction of the field stars was
formed in the so-called clustered mode (i.e.\ in clusters or associations) and
in the dispersed mode. This will in turn provide constraints on molecular
cloud lifetimes and the star formation process.

If the clustered mode of star formation is the dominant one then the resulting
star clusters should dissolve or at least lose a significant fraction of their
stars in order to explain the origin of the field star population. Using
Hipparcos data we can now study in detail several examples of the cluster
dissolution processes in the Solar neighbourhood, which is the subject of
Sect.~\ref{sec:dissolution}

The last link between stellar clusters and field stars is formed by the
so-called moving groups. These are the remnants of old clusters just prior to
complete disintegration and the stars are strung out over large volumes all
along the orbit of the parent cluster around the Galactic centre. If the Sun
happens to be near the orbit of the parent cluster its stars will appear all
over the sky but we may still recognise them as a moving group by their shared
space velocity. These moving groups are important for tracing the early
evolution of the Galactic disk (that is at times before those which can be
probed with open clusters, $\ga10^9$~yr) and for understanding how the
Galactic phase-space distribution, $F({\mathbf x},{\mathbf v})$, comes
about. Studies of the local velocity distribution in terms of moving groups
are discussed in Sect.~\ref{sec:moving} Open questions and future prospects
are summarised in Sect.~\ref{sec:future} Finally, in this review I concentrate
on the observational aspects of the evolution from clusters to the
field. Theoretical discussions can be found in, eg., Binney \& Tremaine (1987)
and Meylan \& Heggie (1997).

\section{Dissolution of Open Clusters and OB Associations Near the Sun}
\label{sec:dissolution}

Mass loss from a cluster proceeds most rapidly during the time of its
formation. In stellar clusters that are large enough massive stars are likely
to form and their stellar winds and radiation will result in the rapid
expulsion of the gas in which the cluster is still embedded. The fraction of
stars that gets unbound depends on the star formation efficiency, the
dynamical state of the stars at the time of gas expulsion, and on the details
of the gas removal process. Generally speaking, if the star formation
efficiency is less than 50\% rapid removal of the gas (with respect to the
cluster crossing time) will result in an unbound and expanding OB
association. For larger star formation efficiencies and slow gas removal a
significant fraction of the stars may remain bound in a cluster (see eg.,
Lada, Margulis, \& Dearborn 1984, and Verschueren \& David 1989). The precise
details of this process are still not understood but the problem can now be
addressed by simulations employing sophisticated N-body models of clusters
embedded in gas as described elsewhere in this volume by Kroupa.

After a bound cluster emerges from the gas it will continue to lose stars on
much longer time scales. Due to stellar dynamical processes stars in the
cluster will wander through phase space and sometimes reach the portions
thereof corresponding to unbound orbits. Many weak encounters between stars
will lead to a gradual `evaporation' of stars from the cluster, a process
which is enhanced by the presence of the Galactic tidal field. Single close
encounters between stars may lead to a large enough velocity change for one of
them to escape the cluster. In this process binaries can be an especially
efficient source of energy, enough to effect the complete disruption of small
($N\la100$) clusters. The remaining processes that may lead to cluster
disruption are encounters with molecular clouds (for open clusters) and
disk/bulge shocking (for globular clusters). Details can be found in, eg.,
Binney \& Tremaine (1987) and Meylan \& Heggie (1997).

\subsection{OB Associations}
\label{sec:associations}

The OB associations near the Sun can be characterised as young ($\la\,$50~Myr)
and loose stellar groupings containing a significant population of B
stars. Their large physical sizes ($\sim\,$10--100~pc) imply very low stellar
mass densities of the order of $0.1$ ${\rm M}_\odot$~pc$^{-3}$, assuming a
`normal' initial mass function such as the one suggested by Scalo (1998). This
assumption is justified by the work of Preibisch \& Zinnecker (1999) who have
shown that the Sco~OB2 association shows no lack of low-mass stars. At these
densities the associations are unstable against Galactic tidal forces (Bok
1934). Hence these groupings are presumably the remnants of clusters in which
stars formed with low efficiency followed by rapid loss of the gas.

De Zeeuw et~al.\ (1999) used Hipparcos proper motions and parallaxes to
accurately establish the stellar content of 12 nearby OB associations down to
mid-F spectral type. The closest and best studied of these associations is Sco
OB2, which consists of three subgroups; Upper Scorpius (US, near the Ophiuchus
star forming region, at a distance of $145\!\pm\!2$~pc), Upper Centaurus Lupus
(UCL, $140\!\pm\!2$~pc) and Lower Centaurus Crux (LCC, $118\!\pm\!2$~pc). More
details can be found in de Zeeuw et~al.\ (1999).

From the distribution of the stars on the sky (see Fig.~9 in de Zeeuw et~al.)
it is clear that the youngest subgroup, Upper Scorpius is much more
concentrated than the two older subgroups. Moreover the sizes of the subgroups
(30, 65, and 45~pc respectively for US, UCL, and LCC) are correlated with
their ages (5, 13, and 10~Myr), which can also bee seen in the case of
Orion~OB1 (Brown, de Geus, \& de Zeeuw 1994). This is consistent with OB
associations being in a state of expansion. In fact the expansion has been
measured directly for Upper Scorpius by Blaauw (1978, see also Blaauw 1991)
and for a number of other associations by various authors (see references in
Brown et~al.\ 1999). The association in the most advanced state of
disintegration that is still recognised as an entity is Cas--Tau, shown in
Fig.~\ref{fig:castau}.

In principle precise proper motions in combination with radial velocities can
be used to trace back the motions of the stars in an expanding association to
their approximate points of origin. This will allow independent estimates of
the ages of associations as well as constraining their configurations very
shortly after they were formed. In this way one can investigate whether a
given association or subgroup consists of merged smaller subunits. Suggestive
evidence for this was found by examining the 3D position of the stars in Upper
Centaurus Lupus (De Bruijne 1999). Constraining the clustering properties of
stars right after their formation process has finished will teach us more
about what modes of star formation occur and how important they are.

\subsection{Open Clusters: the Hyades}
\label{sec:hyades}

The Hipparcos data have stimulated many studies of the nearby open clusters,
with most of the attention going to the Pleiades and Hyades (eg., Robichon
et~al.\ 1999, Pinsonneault et~al.\ 1998, Perryman et~al.\ 1998). The most
detailed studies can be carried out for the Hyades, being the nearest
moderately rich open cluster at only 46~pc from the Sun (Perryman et~al.\
1998). A detailed investigation into its three-dimensional structure and
kinematics was carried out by Perryman et~al.\ (1998). The velocity field was
subsequently investigated more thoroughly by Lindegren, Madsen, \& Dravins
(2000).

Perryman et~al.\ (1998) found that a large fraction of the 218 Hipparcos
members lie outside the central concentration of the cluster. The dynamical
limit of the cluster is roughly indicated by the so-called tidal radius. This
is the distance at which the equipotential cluster surface becomes open, due
to the effects of the Galactic tidal field. For the Hyades the tidal radius is
$\sim\,$10~pc. About 45 stars are located between 10--20~pc from the cluster
centre. This number is consistent with the simulations of Terlevich (1987),
who consistently finds a halo formed by 50--80 stars in the region between
1--2 tidal radii -- some of these stars, despite having energies larger than
that corresponding to the Jacobi limit, are still linked to the cluster after
300--400~Myr. This can be explained by the location of the openings of the
equipotential surface on the $x$-axis (pointing from the Sun towards the
Galactic centre, the Hyades being located at $\ell=180^\circ$); stars can
spend some considerable time within the cluster before they find the windows
on the surface to escape through. Alternatively, stars beyond the tidal limit
can remain bound to the cluster by an angular momentum-like non-classical
integral of motion (eg., H\'enon 1970).

Furthermore the cluster has a prolate shape with axis ratios $1.6$:$1.2$:1,
where the major axis makes an angle of only $\sim\!16^\circ$ with the
direction towards the Galactic centre. This suggests that the cluster is
primarily extended along this direction, although it is possibly also slightly
compressed perpendicular to the Galactic plane. This is consistent with the
extension being caused by stars slowly escaping through the Lagrangian points
on the $x$-axis. The flattening of the equipotential surfaces, perpendicular
to the Galactic plane is clearly evident in the $N$-body simulations by
Terlevich (1987, Fig.~8).

All these observational facts together are clear manifestations of ongoing
dynamical processes in the Hyades cluster. These will lead to the main cluster
being surrounded by a moving group of which Eggen (1982) has already
identified specific candidate members. Ultimately the Hyades cluster as a
whole will turn into a moving group and dissolve into the Galactic field.

\subsection{$\alpha$~Persei and the Cas--Tau association}
\label{sec:castau}

In order to better understand the fate of the Orion Nebula Cluster, Kroupa,
Aarseth, \& Hurley (2000) have done detailed numerical simulations taking into
account both the process of gas expulsion from the cluster and the effects of
the Galactic tidal field. They find that even if the cluster expels 2/3 of its
original mass (i.e.\ the star formation efficiency is 30\%) a bound component
may still survive as an open cluster. This result is rather different from
what was traditionally found by, eg., Lada et~al.\ (1984) and the authors
attribute this to the fact that they take into account gravitational
interactions between neighbours in the radially expanding flow of stars
(details in Kroupa et~al.\ 2000).

The most interesting result in the context of this review is that Kroupa
et~al.\ (2000) find that the cluster remnant looks a lot like the Pleiades
open cluster as measured from properties such as the number of stars and the
distribution of binaries. Hence, they predict that the Orion Nebula Cluster
will eventually evolve into an open cluster surrounded by an expanding OB
association. The expansion velocity they predict for the fastest moving stars
is 2--3~km/s, implying that they will be spread over large a volume (radius of
100--150~pc after 50~Myr) in a short time. Do such systems exist in the Solar
neighbourhood?

In their study of the nearby OB associations de Zeeuw et~al.\ (1999) point out
a possible connection between the old Cas--Tau OB association and the
$\alpha$~Persei cluster. Cas--Tau is the most dispersed and oldest association
near the Sun. At about 190~pc distance, it covers about 100$\times$60 degrees
on the sky and Blaauw (1956) derived an expansion age of $\sim\,$50~Myr. The
Hipparcos Catalogue contains 83 members, all B-type stars (de Zeeuw et~al.\
1999). Nothing is known about the stellar content of Cas--Tau beyond this
spectral type. $\alpha$~Persei is a well-studied open cluster for which
members have been found down to $V\sim21$ (eg., Prosser 1994). De Zeeuw
et~al.\ (1999) established a distance for this cluster based on Hipparcos
parallaxes of $177\!\pm\!4$~pc and its age was most recently determined to be
$90\!\pm\!10$~Myr from the Lithium depletion boundary (Stauffer et~al.\ 1999).

\begin{figure}
\begin{center}
\epsfig{file=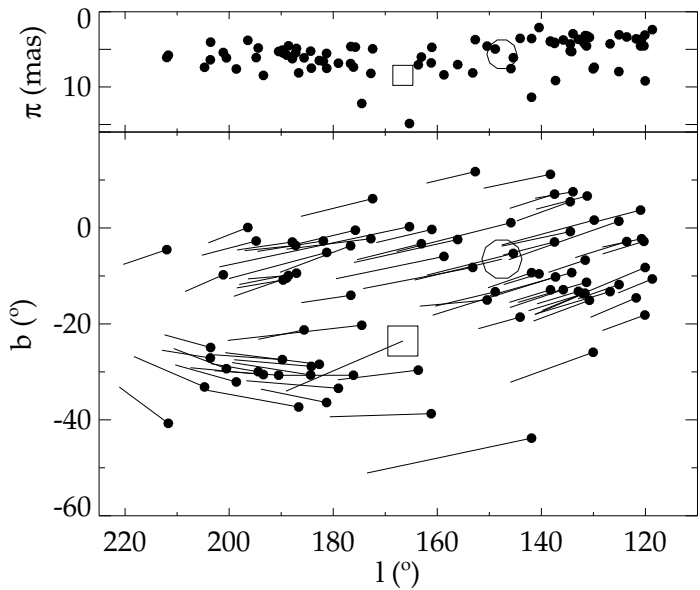,height=3truein}
\end{center}
\caption{Positions and proper motions, in Galactic coordinates, for the
Hipparcos selected members of the Cas--Tau association. The largest proper
motion vector is 75 milli-arcseconds/yr. The upper panel shows the parallaxes
(in milli-arcseconds) as a function of Galactic longitude. Also indicated are
the position and proper motion of the Pleiades (square) and $\alpha$~Persei
(circle). Note how $\alpha$~Per fits in the overall proper motion pattern for
Cas--Tau.}\label{fig:castau}
\end{figure}

Both stellar groupings show very similar kinematics and are also located at
roughly the same distance, which is illustrated in Fig.~\ref{fig:castau}. This
may thus be an example of an association with an open cluster at its core, as
predicted by Kroupa et~al.\ (2000) from their simulations. This suggestion is
further strengthened by the finding by de Zeeuw et~al.\ (1999) that a `halo'
spanning 10$^\circ$ on the sky and consisting mainly of A-stars surrounds the
core of the $\alpha$~Persei cluster. This halo could be the innermost part of
the expanding association surrounding the cluster or it could consist of stars
that did not manage to escape entirely from the cluster after gas
expulsion. The age of Cas--Tau has been determined from the expansion of the
association but it was shown by Brown, Dekker, \& de Zeeuw (1997) that this
will lead to underestimates of the true age. Also, the main sequence turn-off
age for $\alpha$~Persei is $\sim\,$50~Myr (Meynet, Mermilliod, \& Maeder
1993). Hence the ages of both groups of stars could well be the same.

Assuming Cas--Tau has a mass function such as suggested by Scalo (1998) the
number of stars in the mass range $0.1$--1~M$_\odot$ in this dispersed group
of stars could be a few thousand. This association lies at the heart of
Gould's Belt (Fig.~29 in de Zeeuw et~al.\ 1999) and in the same region are
located the Pleiades cluster and a new association found by de Zeeuw et~al.\
(1999); Cep~OB6. If the Pleiades are also accompanied by a surrounding
dispersed population and if Cep~OB6 has a normal mass function, many more
low-mass stars can be expected in this area. This means that the volume of
Gould's Belt could be filled with relatively young (50--100~Myr) low-mass
stars. This could explain the dispersed X-ray population found by ROSAT
(Sterzik et~al.\ 1995).

Speculating a bit further; Kroupa et~al.\ (2000) note that in their
simulations sometimes multiple clusters form in the expanding flow of
stars, which merge later on with the dominating cluster (if they are not
evaporated first by two-body relaxation). This raises the question whether
significant clumping may occur in the parts of the flow that will eventually
end up as the association surrounding the open cluster. Such surviving clumps
could be the progenitors of the recently discovered nearby groupings of
T-Tauri stars, such as the TW~Hya (Webb et~al.\ 1999) and $\eta$~Chamaeleontis
(Mamajek, Lawson, \& Feigelson 1999) groups.

\section{Old Moving Groups and Phase Space Structure Near the Sun}
\label{sec:moving}

The availability of Hipparcos data for the first time allows the analysis of
space velocities (for which parallaxes are needed) for large samples of stars
near the Sun. This has inspired much new research into the question of moving
groups through studies of the distribution of stars in velocity
space. However, in spite of the unprecedented quality of the Hipparcos
astrometric data, studies of the local velocity distribution are still
hampered by the lack of complementary high-precision and homogeneous radial
velocities. Moreover, the Hipparcos Catalogue contains numerous selection
biases which are difficult to characterise, and the conversion from parallaxes
and proper motions to distances and space velocities will introduce further
biases (see Brown et~al.\ 1997).

In this light the most careful study to date of the distribution of nearby
stars in velocity space, $f({\mathbf v})=F({\mathbf x}_\odot,{\mathbf v})$,
was carried out by Dehnen (1998). He used a kinematically unbiased sample of
stars from the Hipparcos Catalogue which have parallax errors of less than
10\%. Assuming that $f({\mathbf v})$ is independent of position on the sky one
can make use of the fact that in different directions one is presented with
varying projections of $f({\mathbf v})$. Using the Fourier slice theorem one
can then invert these projections in order to obtain $f({\mathbf v})$. The
strength of this approach is that radial velocities are not required (see
below).

Dehnen divided his sample into to four subsamples of stars of increasingly red
\bv colour and thus of increasing average age. Apart from identifying the
well-known moving groups he found the following: going from blue to red stars:
the fraction of stars inside maxima in $f({\mathbf v})$, i.e.\ moving groups,
decreases. The number of features in $f({\mathbf v})$ increases, implying the
presence of old moving groups. There is a correlation in the sense that the
older a moving group is, the smaller is its (local) rotational velocity around
the Galaxy. This together with the fact that the width of the distribution of
moving groups increases in $(v_x,v_y)$ but not $v_z$ implies that their
distribution obeys an asymmetric drift relation (the velocity components
$(v_x,v_y,v_z)$ denote motions towards the Galactic centre, in the direction
of Galactic rotation, and towards the North pole).

According to the standard picture of moving groups as remnants of star
clusters one would expect them to be narrow in $v_y$ but this is not the case
for the moving groups near the Sun. In fact they obey an asymmetric drift
relation, which implies that the moving groups are on non-circular orbits.
Star clusters form from molecular clouds which are themselves on circular
orbits. So the orbits of the corresponding moving groups must be shifted to
more eccentric ones. This can only be done by a smooth non-axisymmetric force
field. Normal scattering processes (operating on the background of stars
outside moving groups) would destroy them. An obvious candidate for providing
such a force field is the Galactic bar. The stars in the moving group could be
trapped into a resonance with the force field of the bar and when the force
field slowly changes the resonances shift in orbit space and with them the
trapped stars.

In a follow up study Dehnen (2000) investigated the effect of the galactic bar
on $f({\mathbf v})$ by performing simulations of the velocity distribution in
the outer parts of an exponential disk with a nearly flat rotation curve and a
rotating central bar. The outer Lindblad resonance of the bar indeed causes a
distinct feature in $f({\mathbf v})$ in the form of a bi-modality between a
dominant mode of low-velocity stars centered on the Local Standard of Rest and
a secondary mode of stars mostly moving outward and rotating more slowly than
the Local Standard of Rest. This is exactly what one can see in the
distribution of $(v_x,v_y)$ presented in Dehnen (1998) and indeed in the case
of our galaxy, the outer Lindblad resonance is believed to be near the Solar
radius. There are also other, smaller, features in the simulated velocity
distribution that can be identified with observed features in $f({\mathbf
v})$, such as a ripple caused by orbits trapped in a particular resonance.

Two important conclusions can be drawn from this work. The standard picture of
moving groups on circular orbits, clumped in $v_y$ is too simple. Old moving
groups can be on non-circular orbits and hence be visiting the Solar
neighbourhood from much further inside the Solar circle. Secondly, the stellar
dynamical effect due to the Galactic bar may cause both large and small scale
features in the local $f({\mathbf v})$ which means that one should be careful
when interpreting this velocity distribution in terms of moving groups.

Other searches for moving groups have made use of both Hipparcos data and
radial velocities. This has the advantage of providing a directly measured
$f({\mathbf v})$. However, there are two important disadvantages; only a minor
subsample of all Hipparcos stars have measured radial velocities and these are
inhomogeneous and often imprecise. Secondly, using only stars with known
radial velocities will introduce extra selection effects which are difficult
to accurately characterise.

The effect of the quality of the radial velocity data on the errors in
$(v_x,v_y,v_z)$ is well illustrated by the work of Perryman et~al.\ (1998) on
the Hyades. For this cluster one expects a velocity dispersion of
$\sim0.2$~km/s (based on its mass and assuming virial equilibrium), which is
indeed borne out by the results of Lindegren et~al.\ (2000) which are based on
Hipparcos astrometry only. The results of Perryman et~al.\ (1998) show that
even when using the best available (precise and homogeneous) radial velocities
for known single stars in the Hyades, the measured velocity dispersions are of
the order of $0.5$--$1.3$~km/s, where $0.2$~km/s is due to the astrometric
errors alone. Using all available radial velocities for the Hyades they
measure velocity dispersions of $0.6$--$2.8$~km/s. These are entirely due to
errors caused by the inhomogeneity and varying quality of the radial velocity
data as well as the presence of undetected binaries. Hence any interpretation
of the local velocity distribution should be restricted to structures well
above the 3~km/s level. A case in point is the study by Chereul, Cr\'ez\'e, \&
Bienaym\'e (1998) who claim, on the basis of a wavelet analysis of $f({\mathbf
v})$, to find substructures in moving groups at the level of $3.8$ and
$2.4$~km/s.

Finally, because the velocities of true moving groups overlap with the
velocity distribution of the background stars it is important to employ age
and metallicity discrimination in order to definitively establish their
reality. This is also needed in order to interpret the birth places of the
stars in the moving groups in terms of the metallicity distribution and
evolution of the Galactic disk. Examples are the work by Feltzing \& Holmberg
(2000) on the HR1614 moving group (see also Holmberg in this volume) and
Asiain et~al.\ (1999).

\section{Open Issues and the Way Forward}
\label{sec:future}

In order to make serious progress in the understanding of the Solar vicinity
and the origin of the field star population at least the following questions
should be answered: What is the detailed space and age distribution of the
dispersed X-ray population in Gould's Belt? Does the observed inclination in
the distribution of young stars require some special explanation? That is, is
Gould's Belt with its population of young stars and clusters a typical large
scale star formation region in the Galaxy? Can its evolution serve as a model
for the origin of field stars? Answering these questions will require more
simulations, a more complete stellar census of Gould's Belt, including
parallaxes, proper motions and radial velocities, and a settling of the
question of the ages of the relevant stellar populations (see Favata et~al.\
1998, for a discussion of age indicators for the dispersed X-ray population).

When it comes to understanding the old moving groups and their evolution we
need to answer the following questions: How much structure in the local
velocity distribution is simply due to large scale dynamical effects operating
on stars in the Galactic disk? How do we separate the true moving groups from
the structured background? To answer the latter question one should take ages
and metallicities of stars into account and preferably analyse the local phase
space distribution in terms of quantities that are (approximately) conserved
along stellar orbits, such as angular momentum and energy.

These questions can only be answered by direct measurements of phase space
(positions and velocities) for a sample of stars over a large (several kpc)
volume around the Sun. The sample should be well defined and the data
homogeneous. Finally we need detailed information of the internal phase space
structure of the OB associations and open clusters located throughout this
volume. This will enable a reconstruction of the clustering properties of the
stars very shortly after their formation.

Such requirements translate into a desire for high quality proper motions and
parallaxes, at a precision level of 10 to 100 micro-arcseconds. These should
be complemented by multi-colour intermediate band photometric data, which are
needed in order to carry out age and metallicity discrimination. Finally,
homogeneous and precise (at the few km/s level) radial velocities are also
needed to round out the measurements of phase space.

For such a programme the future looks very bright indeed, given the recent
approval of funding for several proposed space-astrometry missions, which are
listed in Table~\ref{tab:missions}. Together the FAME (Horner et~al.\ 1999)
and DIVA (R\"oser 1999) missions will already provide tremendous improvements
in our knowledge of the local phase space distribution. However, neither of
these missions will provide intermediate band photometry or radial velocities
so a dedicated effort from the ground would be needed in combination with
clever analysis techniques relying on the astrometric data alone. Ultimately
GAIA (Gilmore et~al.\ 2000) will provide both the required photometry and
radial velocities. Also, because of the much larger reach of GAIA in distance
we can apply what we will have learned from the local phase space distribution
to a detailed understanding of the field star population and formation history
of the entire Galaxy.

\begin{table}[htb]
\caption[tab:missions]{Future survey-type space-astrometry missions. The
columns list the responsible agency, the scheduled launch date, the number of
stars to be observed, the magnitude limit, the measurement precision (in
milli-arcseconds) achieved at varying magnitudes, and the complementary data
gathered. The Hipparcos mission is shown for comparison. Note that the GAIA
data will be complete to $V=20$.}
\label{tab:missions}
\begin{center}
\leavevmode
\footnotesize
\begin{tabular}{lrrrrrrl}
\hline \\[-8pt]
Mission & Agency & Launch & $N_{\rm stars}$ & $V_{\rm lim}$ &
\multicolumn{2}{c}{Precision} & Compl.\ data\\
 & & & & & (mas) & (mag) & \\[2pt]
\hline \\[-8pt]
Hipparcos & ESA & 1989 & $1.2\times10^5$ & 12 & 1 & 10 & $B$, $V$ phot. \\
DIVA & DLR & 2003 & $4\times10^7$ & 15 & $0.2$ & 9 & spectrophot. \\
     &     &      &               &    &   5 & 15 & \\
FAME & USNO/ & 2004 & $4\times10^7$ & 15 & $0.05$ & 9 & 4 Sloan bands\\
     & NASA  &      &               &    & $0.3$ & 15 & \\
GAIA & ESA & 2012 & $10^9$ & 20 & $0.003$ & 12 & 4 broad bands \\
     &     &      &        &    & $0.01$  & 15 & 11 medium bands\\
     &     &      &        &    & $0.2$   & 20 & radial velocities\\
\hline \\

\end{tabular}
\end{center}
\end{table}

\end{document}